\documentclass[twocolumn,showpacs,preprintnumbers,amsmath,amssymb,superscriptaddress,prb]{revtex4}

\usepackage{graphicx}
\usepackage{dcolumn}
\usepackage{bm}

\newcommand{\rmi}{i}

\newcommand{\mysection}[1]{{\it #1}.---}

\usepackage{amsmath}	
\usepackage{txfonts,wasysym,ulem,url}
\usepackage[usenames]{color}

\begin{document}


\title{
Chiral condensate with topological degeneracy in  graphene 
and its manifestation in edge states
}

\author{Yuji Hamamoto}
 \affiliation{%
Institute of Physics, University of Tsukuba, Tsukuba 305-8571, Japan%
}%

\author{Hideo Aoki}
 \affiliation{%
Department of Physics, University of Tokyo,  Hongo, Tokyo 113-0033, Japan%
}%

\author{Yasuhiro Hatsugai}%
\email{hatsugai@sakura.cc.tsukuba.ac.jp}
 \affiliation{%
Institute of Physics, University of Tsukuba, Tsukuba 305-8571, Japan%
}%
 \affiliation{%
Tsukuba Research Center for Interdisciplinary Material Science, University of Tsukuba, Tsukuba 305-8571, Japan%
}%

\date{\today}

\begin{abstract}
Role of chiral symmetry in many-body states
of
graphene in strong magnetic fields is theoretically studied with the honeycomb lattice model.  
For a spin-split Landau level 
where the leading electron-electron interaction is 
the nearest-neighbor repulsion,
a chiral condensate 
is shown to be, within the subspace of $n=0$ Landau level, an 
exact many-body ground state with 
a finite gap, for which calculation of Chern numbers reveals that
the ground state is a Hall insulator with a {\it topological degeneracy} 
of two.  
The topological nature of the ground state is shown to manifest itself as 
a Kekul\'ean bond order along armchair edges,
while the pattern melts in the bulk due to
quantum fluctuations.  
The whole story can be regarded as a realization of the 
bulk-edge correspondence peculiar to the chiral symmetry.
\end{abstract}

\maketitle

\mysection{Introduction}%
While the physics of graphene started from the one-body 
electronic structure as a Dirac fermion, 
a possible relevance of electron correlation in graphene has 
been intensively studied after a gap opening in the $n=0$ 
Landau level (LL)
was experimentally observed in strong magnetic fields.~\cite{PhysRevLett.96.136806,PhysRevLett.99.106802}
Since it is difficult to explain the gap within a simple one-body problem,
considerable theoretical efforts have ensued to clarify many-body effects 
in graphene quantum Hall (QH)
systems.~\cite{PhysRevLett.96.256602,
PhysRevB.74.075422,
PhysRevB.74.161407,
PhysRevB.74.195429,
PhysRevB.75.165411,
2007SSCom.143..504A,
PhysRevLett.99.196802}
However, little attention has been paid on how  
many-body effects should reflect the {\it chiral symmetry} 
in the graphene QH regime, which is after all 
a fundamental symmetry inherent in graphene's honeycomb lattice.
On the one-body level, the effects of chiral symmetry in graphene is well 
understood: 
To start with, the symmetry guarantees the 
emergence of doubled Dirac cones in the Brillouin zone, 
which can be interpreted as a two-dimensional analogue
of the Nielsen-Ninomiya theorem well-known 
in the four-dimensional lattice gauge theory.  
We can even examine the wave functions 
in terms of Aharonov-Casher argument, which states that
chiral symmetry {\it topologically protects}
the degeneracy of the $n=0$ LL against random gauge fields.~\cite{PhysRevA.19.2461}
A similar situation occurs for ripples in a graphene sheet, 
which can be modelled by random hopping amplitudes.  
Kawarabayashi {\it et al.} have shown that  the $n=0$ LL exhibits
an anomalously sharp (delta-function-like) density of 
states (DOS) as soon as the spatial wavelength of the ripple 
exceed a few lattice constants.~\cite{PhysRevLett.103.156804}

In the presence of electron-electron interactions, 
on the other hand, 
the role of chiral symmetry has primarily been investigated
in zero magnetic fields 
in the context of spontaneous symmetry breaking.~\cite{PhysRevB.74.195429,
PhysRevLett.102.026802,
PhysRevB.79.165425,
PhysRevB.82.121403}
While these studies mainly employ a Dirac field model 
in a continuum space to discuss many-body gap formation, 
such an effective treatment may well overlook 
the essence of graphene's chiral symmetry,
which is intimately related to the underlying honeycomb lattice.

With this background, we shed light in the present work on 
how the chiral symmetry influences the many-body problem
in graphene QH effect, by
fully taking account of the lattice structure.  
We first examine 
the many-body problem with exact diagonalization
in a subspace projected onto the $n=0$ LL.
Working on the subspace enables us to classify 
many-body states according to a notion of the {\it total chirality} of 
the filled zero modes.  In terms of this, 
for a ``bipartite'' electron-electron interaction such as
the nearest neighbor repulsion,
which is the dominant interaction for a spin-split LL, 
we show that the many-body ground state is {\it exactly} identified to be 
a chiral condensate with a topological degeneracy of two.  
We confirm numerically that there exists a finite energy 
gap to the first-excited state, 
which makes the Chern number of the ground state  well-defined.
The total Chern number contributed by the filled zero modes
along with the negative energy states (``Dirac sea") 
turns out to be zero, 
which implies the system is a Hall insulator with vanishing Hall conductance.

Despite the cancellation of the Chern number in the bulk, however, 
we move on to show that the topological nature of the chiral condensate 
is in fact made manifest as an emergence of a Kekul\'ean bond order in the edge state 
along armchair edges of the honeycomb lattice 
(in sharp contrast to zigzag edge states in one-body problem).  
On the mean-field (MF)  level 
the Kekul\'e pattern is shown to appear in the bulk 
with a  Kekul\'ean degeneracy of three, 
so that the present result amounts that the mean-field 
order, while dissolved in the topologically-degenerate 
chiral condensate in the bulk, resurfaces along an armchair edge.  
This can be interpreted as an example of bulk-edge correspondence,~\cite{PhysRevLett.71.3697}
which states that a topologically nontrivial bulk state
should always accompany a characteristic edge state.
In the case of graphene, the bulk-edge correspondence 
becomes peculiar in that the chiral condensate 
manifests itself as a freezing of a Kekul\'ean bond order along a 
specific (armchair) edges.

\mysection{Chiral symmetry}%
To model interacting electrons on a honeycomb lattice,
we assume that spin degeneracy is lifted by a Zeeman splitting, 
so that the leading Coulomb interaction reduces to the 
nearest-neighbor repulsion.
The 
Hamiltonian then reads $\mathcal{H}=\mathcal{H}_{\rm kin}+\mathcal{H}_{\rm int}$, where 
the kinetic term
$\mathcal{H}_{\rm kin}=-t\sum_{\langle ij\rangle}
(e^{\rmi\theta_{ij}}c^\dagger_{i}c_{j}+{\rm H.c.})
\equiv c^{\dagger}H_{\rm kin}c$
describes electron hopping between adjacent sites
$\langle ij\rangle$ with strength $t>0$.
The magnetic field is included as the Peierls phase
$\theta_{ij}$ such that magnetic flux per elementary hexagon equals
$\sum_{\hexagon}\theta_{ij}=2\pi\phi$ in units of a magnetic flux quantum $h/e$.
For a honeycomb lattice with $N_{\bullet(\circ)}$
sites in sublattice $\bullet(\circ)$,
$c^{\dagger}=(c^{\dagger}_{\bullet},c^{\dagger}_{\circ})$
with $c^{\dagger}_{\bullet(\circ)}$ a row of creation operators
for sublattice $\bullet(\circ)$
and $H_{\rm kin}$ is a square matrix of dimension $N_{\bullet}+N_{\circ}.$
If we introduce $\Gamma={\rm diag}(I_{\bullet},-I_{\circ})$
with an identity matrix $I_{\bullet(\circ)}$ of dimension $N_{\bullet(\circ)}$,
the kinetic term satisfies an anticommutation relation
$\{H_{\rm kin},\Gamma\}=0$, 
which defines the chiral symmetry.
The symmetry implies that, 
if $\psi_{k}$ is the $k$-th eigenvector with energy $\varepsilon_{k}$,
a chiral partner
$\Gamma\psi_{k}$ exists with an energy $-\varepsilon_{k}$.
This makes the $n=0$ LL special
in that we can take
$\psi_{k}$ as an eigenstate of $\Gamma$ 
as $\Gamma\psi_{k\pm}=\pm\psi_{k\pm}$.
The interaction between spin-polarized electrons 
is expressed in a particle-hole symmetric form as
$\mathcal{H}_{\rm int}=\sum_{i\ne j}V_{ij}
\left(n_i-\frac{1}{2}\right)\left(n_j-\frac{1}{2}\right)
=\frac{1}{2}\sum_{i\ne j}V_{ij}
(c^\dagger_ic^{\dagger}_{j}c_jc_i
+c_ic_{j}c^{\dagger}_jc^{\dagger}_i)+{\rm const.}$,
where $V_{ij}$ is the strength of electron-electron interaction,
$n_i\equiv c^\dagger_ic_i$ the number operator at site $i$.

\mysection{Chiral condensate}%
We start with an investigation of 
the many-body problem
at half filling.
Since it is difficult to treat all the many-body states exactly,
we shrink the Hilbert space
by projecting onto the $n=0$ LL.
Such a treatment is valid as long as $|V_{ij}|$ is perturbatively small
compared with the Landau gaps around the $n=0$ LL.~\cite{detail}
In the $n=0$ LL,
we take a zero mode multiplet $\psi=(\psi_{+},\psi_{-})$
where we have decomposed it into 
eigenstates of the chiral operator, 
$\psi_{\pm}=(\psi_{1\pm},\cdots,\psi_{M_{\pm}\pm})$ with degeneracy $M_{\pm}$.
Note that the zero modes are localized on each of the sublattices
as $\psi_{+}=\frac{1}{\sqrt{2}}\left(
\scriptsize{
\begin{array}{c}
\psi_{\bullet}\\
0
\end{array}
}
\right)$ and $\psi_{+}=\frac{1}{\sqrt{2}}\left(
\scriptsize{
\begin{array}{c}
0\\
\psi_{\circ}
\end{array}
}
\right)$.
If we introduce the negative-energy multiplet $\varphi=(\varphi_{1},\varphi_{2},\cdots)$
such that $H_{\rm kin}\varphi_{k}=\varepsilon_{k}\varphi_{k}$
with $\varepsilon_{k}<0$,
$(\psi_{+},\psi_{-},\varphi, \Gamma\varphi)$ form a complete set,
so that the fermion operator is expanded as
$c=\psi d+\varphi d_{<}+\Gamma\varphi d_{>}$,
where
$d=\left(
\scriptsize{
\begin{array}{c}
d_{+}\\
d_{-}
\end{array}
}
\right)$
and $d_{\lessgtr}$ are columns of fermion operators
for the zero modes and the $\varepsilon\lessgtr0$ states, respectively.
In the projected subspace,
the total Hamiltonian is written, up to a constant, as
$\tilde{\mathcal{H}}=\frac{1}{2}\sum_{i\ne j}V_{ij}
(\tilde{c}^{\dagger}_{i}\tilde{c}^{\dagger}_{j}\tilde{c}_{j}\tilde{c}_{j}
+\tilde{c}_{i}\tilde{c}_{j}\tilde{c}^{\dagger}_{j}\tilde{c}^{\dagger}_{i})$,
where the projected fermion operator, $\tilde{c}_{i}\equiv\psi d$,
no longer satisfies the canonical anticommutation relations.
Still, the generator of the total chirality of the filled zero modes
can be defined as $\mathcal{G}=\tilde{c}^{\dagger}\Gamma\tilde{c}
=d^{\dagger}_{+}d_{+}-d^{\dagger}_{-}d_{-}$.
Due to the invariance of $\tilde{\mathcal{H}}$ for the chiral transformation
$\tilde{\mathcal{H}}\mapsto
e^{\rmi\theta\mathcal{G}}\tilde{\mathcal{H}}e^{-\rmi\theta\mathcal{G}}
=\tilde{\mathcal{H}}$,
the total chirality is conserved, $[\tilde{\mathcal{H}},\mathcal{G}]=0$.
Then the many-body states are classified according to the {\it total chirality} 
$\chi_{\rm tot}$.
For repulsive interactions $V_{ij}>0$,
$\tilde{\mathcal{H}}$ is semi-positive definite.
Furthermore, when the interactions are bipartite (i.e., only 
act between different sublattices, $V_{i\in\bullet,j\in\bullet}=V_{i\in\circ,j\in\circ}=0$) as the case with the nearest-neighbor interaction, 
chiral condensates $|G_{\pm}\rangle=d^{\dagger}_{1\pm}\cdots d^{\dagger}_{M_\pm\pm}
|D_{<}\rangle$ (with $|D_<\rangle\equiv\prod_{m}(c^{\dagger}\varphi)_{m}|0\rangle$ denoting the Dirac sea) 
constitute a {\it ground state doublet} $\Psi=(|G_{+}\rangle,|G_{-}\rangle)$,
since $c_{j}c_{i}|G_{\pm}\rangle=c^{\dagger}_{j}c^{\dagger}_{i}|G_{\pm}\rangle=0$
for $i\in\bullet$ and $j\in\circ$.~\cite{detail,hamamoto}
We call this a topological degeneracy of two.
Unlike a simple charge density wave (CDW),
one may mix $|G_{+}\rangle$ and $|G_{-}\rangle$
through a unitary transformation $\Psi\mapsto\Psi^{\omega}\omega$
with $\omega\in{\rm U}(2)$ even in a finite system.
One can numerically confirm
that the chiral condensate remains the ground state
unless the non-bipartite potential is large,~\cite{PhysRevLett.99.196802}
even though the ground-state energy is nonzero.
In the rest of the paper,
we focus on the leading interaction, i.e., the nearest-neighbor repulsion
for simplicity.

\begin{figure}[t]
\begin{center}
 \includegraphics[scale=1.15]{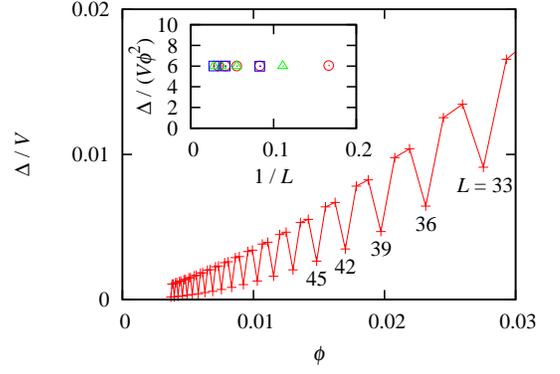}
\caption{\label{fig:phi-dependence}
(color online)
Magnetic-field dependence of the many-body gap $\Delta$ in the $n=0$ LL.
The result is displayed for 30 electrons 
with the consecutive values of the linear dimension of 
the system, $L$, connected by a line with the results 
for $L\equiv 3$ (mod 3) forming a lower envelope 
($\Delta\propto\phi^2$).  
The inset shows the gap at $L=3l$ against $1/L$ 
for the magnetic flux $\phi=1/12$ (circle), $\phi=1/27$ (triangle), 
and $\phi=1/48$ (square).
}
\end{center}
\end{figure}

\mysection{Many-body gap}%
We next calculate excitation energies numerically
with the exact diagonalization method.
In the projected subspace,
the strength of the nearest-neighbor repulsion $V>0$ is the only 
energy scale, which acts as the unit of energy.
Full energy spectra of $\tilde{\mathcal{H}}$ for finite systems
suggest that the first excited state appears in the sector of
$\chi_{\rm tot}=\pm(M_\pm-2)$,
which is created from a chiral condensate $|G_{\pm}\rangle$
by single chirality-flippings 
analogous to the projected single-mode approximation.~\cite{PhysRevLett.54.581,PhysRevB.33.2481,PhysRevLett.73.3568}
Noticing this,
we further shrink the Hilbert space
by focusing on the sector of $\chi_{\rm tot}=\pm(M_\pm-2)$.
This enables us to obtain the energy of the first excited state,
or the energy gap $\Delta$,
with calculation cost of the order of $\mathcal{O}(M_\pm^2)$.
We consider a system on a torus composed of $2L^2$ lattice sites
with a linear dimension $L$.
For investigating a weak-field regime, we adopt the string gauge,~\cite{PhysRevLett.83.2246}
where a magnetic flux is given by $\phi=m/L^2$ with an integer $m$ $(>0)$
and $M_\pm=m$ zero modes are obtained for each chirality.

In Fig.~\ref{fig:phi-dependence},
we plot $\Delta$ for $30$ electrons as a function of $\phi$ with $L$ changed consecutively.  We immediately 
notice that 
the result exhibits a marked periodicity of three, where 
the values for $L\equiv 3$ (mod 3) form a clear lower 
envelope with a scaling $\Delta\propto\phi^2$,
while those for $L\ne3l$ deviates from this.  
The latter behavior is considered to be a finite-size effect, 
since the deviation diminishes with the sample size.  
To confirm the scaling, the inset plots $\Delta/\phi^2$ at $L=3l$ 
against $1/L$ for various values of $\phi$,
which indicates the scaling law $\Delta\propto\phi^2$ is 
very accurately obeyed.

\mysection{Hall conductance}%
Let us now consider Hall conductance of the chiral condensate.
By the Niu-Thouless-Wu formula,~\cite{PhysRevB.31.3372}
the Hall conductance is written with the Chern number~\cite{PhysRevLett.49.405} as
\[
\sigma_{xy}=\frac{e^{2}}{h}\frac{1}{N_{D}}C,\qquad
C=\frac{1}{2\pi\rmi}\int{\rm Tr}_{N_{D}}dA,\qquad
A=\Psi^{\dagger}d\Psi
\]
where $N_{D}$ is the degeneracy
and $A$ is the non-Abelian Berry connection 
that describes multiplets.~\cite{2004JPSJ...73.2604H}
In terms of the basis that diagonalizes $\mathcal{G}$, we have 
$C=C_{+}+C_{-}$ with
$C_{\pm}=\frac{1}{2\pi\rmi}\int\langle dG_{\pm}|dG_{\pm}\rangle$.
Each term is further decomposed as $C_{\pm}=C_{\psi_{\pm}}+C_{D_{<}}$
with $C_{\psi_{\pm}}=\frac{1}{2\pi\rmi}\int{\rm Tr}_{M_{\pm}}
d\psi^{\dagger}_{\pm}d\psi_{\pm}$
and $C_{D_{<}}=\frac{1}{2\pi\rmi}\int{\rm Tr}d\varphi^{\dagger}d\varphi$.
By the charge conjugation,
we have $C_{\psi_{+}}+C_{D_{<}}=-(C_{\psi_{-}}+C_{D_{>}})$
with $C_{D_{>}}
=\frac{1}{2\pi\rmi}\int{\rm Tr}(\Gamma d\varphi)^{\dagger}\Gamma d\varphi=C_{D_{<}}$.
Thus the total  Chern number of the ground-state doublet
vanishes as $C=C_{\psi_{+}}+C_{\psi_{-}}+2C_{D_{<}}=0$,
which may be called a topological cancellation.
This implies that the chiral condensate is a Hall insulator
with a nontrivial topological degeneracy $N_{D}=2$.

\begin{figure}[t]
\includegraphics[scale=1.15]{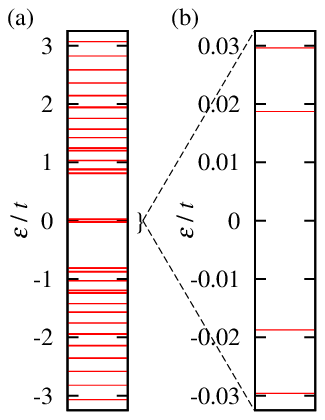}
\hfill
\includegraphics[scale=1.15]{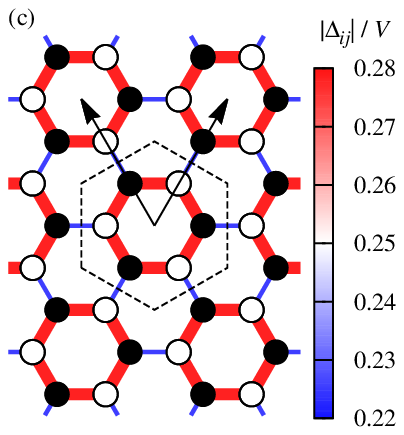}
\caption{\label{fig:bond-order}
(color online) Mean-field results for the energy spectrum (a),
its blowup around the $n=0$ LL (b),
and the bond strength $|\Delta_{ij}|$ plotted in a real-space (c).
The parameters are $L=15,\phi=1/15$, and $V/t=0.25$.
A dashed hexagon in (c) is an enlarged unit cell with arrows primitive vectors.}
\end{figure}

\mysection{Bond order}%
As have been confirmed in various systems,
while topological phases are featureless in a bulk,
they show characteristic boundary states.~\cite{PhysRevLett.71.3697}
So a natural question 
we can pose here is: do the edge states in the present system 
exhibit special features despite the bulk Chern number 
being zero?  
Before presenting the result, however, let us first 
have a look at the mean-field state in the present 
system in the bulk, which will turn out to be instructive.   
One virtue of a mean-field picture~\cite{hatsugai08up} is that 
we can introduce a bond order, 
$\Delta_{ij}\equiv V\langle c^\dagger_ic_j\rangle$, 
for adjacent sites $\langle ij\rangle$.~\cite{PhysRevB.37.3774,
PhysRevB.39.11413,
PhysRevB.45.4027,
PhysRevB.61.16377,
PhysRevLett.98.146801}
The dominant part of the MF Hamiltonian is given by
$\mathcal{H}_{\rm MF}=-\sum_{\langle ij\rangle}[(te^{\rmi\theta_{ij}}
+\Delta_{ij}^\ast)c^\dagger_ic_j+{\rm H.c.}]$,
where $\Delta_{ij}$ is determined self-consistently by diagonalizing
$\mathcal{H}_{\rm MF}$.
A spontaneous symmetry breaking is induced by the many-body effect for weak magnetic fields, 
where the density of states has a sharp peak at  
the Fermi energy.
In Fig.~\ref{fig:bond-order},
we show a typical MF result for the ordered phase.
The energy spectrum is plotted in panel (a),
where the qualitative structure of the LLs is preserved.
This comes from the fact that the convergent order parameters turn out 
to retain the initial Peierls phase
as $\Delta_{ij}=|\Delta_{ij}|e^{-\rmi\theta_{ij}}$.
The influence of the electron-electron interaction appears most prominently in the $n=0$ LL,
where a finite gap of the order of $\phi$ opens as shown in the blowup Fig.~\ref{fig:bond-order}(b).
To see how the symmetry is broken in the mean field, 
we show in panel (c) a real-space image of the 
bond order $|\Delta_{ij}|$, 
which is seen to exhibit a Kekul\'e pattern.~\cite{2008PhyE...40.1530H,hatsugai08up}
This makes the unit cell enlarged, which causes $K$ and $K'$ points 
to be coupled, and this in turn opens a finite gap.  In this sense 
we can regard this a Peierls transition in the honeycomb lattice.

On the other hand,
the chiral condensate with its topological degeneracy of two
exhibits in the bulk no bond order as we have seen in Fig.~\ref{fig:bond-order}.
Due to the quantum fluctuation, the bond order of the mean field
is destroyed and the quantum liquid ground state is realized.

\mysection{Edge and defect states}%
We are now in a position to ask the question: what kind of 
edge states does the chiral condensate accommodate?
Based on the bulk-edge correspondence,
we may expect a non-trivial behavior of the many-body states near edges.
A prime example is the 
fractional
QH states in a 2DEG, where 
a CDW-like behavior emerges along edges
of a
ribbon,~\cite{PhysRevB.50.17199}
while in the bulk it melts into the Laughlin liquid with no long-range order,
which has a $q$-fold degeneracy of the fractional QH states at filling $\nu=1/q$.
Note that a honeycomb lattice with edges
has QH edge states whose mode lies in a LL gap.  
To perform the projection onto the $n=0$ LL we set an energy cutoff,
the choice of which is shown to have little influence on the edge states shown below.

\begin{figure}[t]
\includegraphics[scale=1.15]{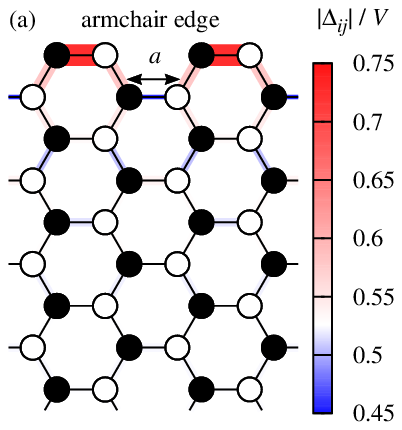}
\hfill
\includegraphics[scale=1.15]{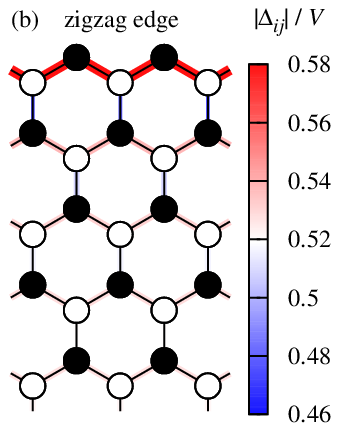}
\caption{\label{fig:bond-chiral}
(color online)
Bond strength near an armchair (left) or zigzag (right) edge of the doubly-degenerate chiral condensate.
Magnetic flux is $\phi=1/192$ for which the magnetic length
is $l_{B}\simeq 8.9a$.
The bond order decays in a bulk away from the edges.
}
\end{figure}

In Fig.~\ref{fig:bond-chiral} we show
$|\Delta_{ij}|$
for the chiral condensate
plotted in a real space near armchair and zigzag edges.
In panel (a),
we can see that a Kekul\'e-type bond order 
reminiscent of the mean-field result in Fig.~\ref{fig:bond-order}
emerges along the armchair edge.
This is the
key result in the present work.
The enhancement in bonds rapidly decays away from the edge
in a few lattice constants,
and $|\Delta_{ij}|$ slightly oscillates
with a length scale of the order of the magnetic length
$l_{B}\sim a/\sqrt{\phi}$
with $a$ being the interatomic spacing.
This may naively seem to be analogous to the
fractional QH edge states in a 2DEG, 
but here the honeycomb lattice structure is essential 
in the ground state.  Indeed,
the ring pattern is locked along the armchair edge 
in a Kekul\'e pattern, 
while this is not the case with zigzag edges [see panel (b)].
In the latter case, 
the ring pattern is blurred by the translational symmetry along a zigzag edge, 
and a very weak stripe pattern parallel to the edge appears.
These patterns related with the three-fold degeneracy of the 
Kekul\'e pattern  are washed out in the bulk chiral condensate.  
All these are a specific property of a honeycomb lattice model.

\begin{figure}[t]
\includegraphics[scale=1.15]{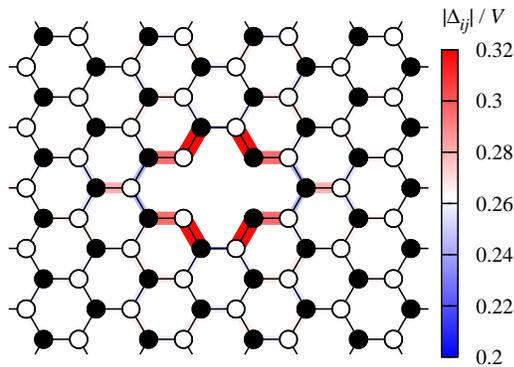}
\caption{\label{fig:bond-defect}
(color online)
Bond strength of the chiral condensate
near a divacancy composed of two adjacent sites missing.
Magnetic flux is $\phi=1/1200$
for which the magnetic length is $l_B\simeq 22.3a$.  
The bond order reflects twofold axial symmetry of the divacancy.
}
\end{figure}

We can further endorse that the lattice structure is at the core 
by looking at the states around lattice defects.
When a single atom is removed from the bulk honeycomb lattice,
one-body localized zero modes appear that are protected
by the chiral symmetry.~\cite{PhysRevLett.89.077002,2009SSCom.149.1061H}
In the presence of the electron-electron interactions, however,
local chiral symmetry breaking occurs spontaneously to lower the energy
by inducing effective hopping in the same sublattice.~\cite{2004PhyE...22..679R,
2009SSCom.149.1061H}
Then what if two point defects come close to each other?
Such a divacancy
consists of two adjacent missing atoms, 
and is recently observed experimentally
in ion-irradiated carbon samples.~\cite{PhysRevB.85.121402}
We expect that the chiral symmetry may be partially recovered
with a reconfiguring of the two symmetry-breaking bonds.
We plot in Fig.~\ref{fig:bond-defect} the 
bond order for the chiral condensate near the divacancy.
We do confirm that
enhancement of the bond order near the divacancy
which can be considered due to the revival of the chiral symmetry.

We have thus shown that the bond order emerges along edges 
and around vacancies, 
despite the topological cancellation $C=0$ 
that might first seem to wipe out any signature of the chiral condensate.
Indeed, the charge density $\langle n_i\rangle$ itself is uniform
for the
chiral condensate even along edges,
which is due to the invariance of the chiral condensate
for the charge conjugation.
Thus it is the bond order $|\Delta_{ij}|$ that we have to 
look at as a probe for the chiral condensate.  
Thus the bond order provides a new probe for the many-body effect
in half-filled graphene applied a magnetic field.
The bond order near
edges
should be observable experimentally with some imaging techniques such as 
Green's function scanning tunneling microscopes.~\cite{PhysRevLett.74.306,PhysRevB.51.5502}
Since the amplitude of
$|\Delta_{ij}|/V=|\langle c^\dagger_ic_j\rangle|$
is of the order of the magnetic flux $\phi$,
the magnetic field should have significant magnitudes.

{\it Summary.}---%
The many-body ground state
at half filling
in the honeycomb lattice is identified as a doubly-degenerate chiral condensate for a spin-split Landau level.  
The many-body effect opens a finite energy gap,
which makes the chiral condensate a generic topological insulator.
However, the system has a peculiar manifestation of 
the bulk-edge correspondence in topological systems 
as an emergence of a bond order with a Kekul\'e pattern 
along armchair edges  in an exact ground state, 
while the pattern is dissolved in the bulk. 

{\it Acknowledgement.}---%
The computation in this work has been done
with the facilities of the Supercomputer Center,
Institute for Solid State Physics, University of Tokyo.
This work was supported in part by Grants-in-Aid
for Scientific Research No. 23340112 and No. 23654128 from the JSPS.


\begin{thebibliography}{37}
\bibitem{detail}
\bibinfo{note}{The detail will be given elsewhere.}
\end{thebibliography}
\end{document}